\documentclass[11pt]{revtex4}
\usepackage[utf8]{inputenc}       
\usepackage{xcolor}               
\usepackage{amsmath,amssymb,amsfonts,latexsym,mathtools}
\usepackage{graphicx}           
\usepackage{float}                
\usepackage{multirow,bigdelim}    
\usepackage{booktabs,array}       
\usepackage{cancel,bigints}      
\usepackage{subfigure}            
\usepackage{bbm}                  
\usepackage{eurosym}            
\usepackage{rotating}  

\begin{document}

\title{Phase Transitions of Electromagnetically Charged Black Holes in Lovelock Gravity
with Nonconstant Curvature Horizons}

     \author{
        N. Farhangkhah$^{1}$\footnote{email address: Ne.Farhangkhah@iau.ac.ir},
        S. Hajkhalili$^{2,3}$\footnote{email address: S.hajkhalili@gmail.com} 
        }
     \affiliation{
        $^1$Department of Physics, Shi.C., Islamic Azad University, Shiraz, Iran\\
        $^2$Department of Physics, School of Science, Shiraz University, Shiraz 71454, Iran \\
        $^3$Biruni Observatory, School of Science, Shiraz University, Shiraz 71454, Iran \\
        }

\begin{abstract}

We present the most general class of charged black hole solutions in third-order Lovelock gravity
in even-dimensional spacetimes, incorporating an electromagnetic field and nonconstant-curvature horizons, 
 which significantly influence the geometry for $n\geq 8$.
Unlike uncharged solutions, the near-origin behavior of the metric exhibits a timelike singularity
for the electrically charged black holes. Thermodynamic stability is analyzed in both grand canonical
 and canonical ensembles. In the grand canonical ensemble, stability, determined by the positivity of 
 both temperature and the Hessian determinant, imposes a lower bound on the event horizon radius below 
 which black holes become unstable; this bound can vanish for specific electric, magnetic, or metric 
 parameters, allowing stable configurations across all horizon sizes. In the canonical ensemble, 
 stability is governed by the heat capacity, revealing both first- and second-order phase transitions. 
 First-order transitions occur when the heat capacity vanishes or diverges at unphysical states, 
 while second-order transitions take place between physical states and exhibit van der Waals–like behavior. 
 Consequently, small and large black holes remain thermodynamically stable, whereas intermediate-sized configurations are unstable.

\end{abstract}

\maketitle

\section{Introduction}

General relativity provides an exceptionally accurate description of the universe
on intermediate and large scales. However, it is widely anticipated that Einstein’s
theory will become inadequate at very short distances or at energy scales approaching the Planck scale.
Motivated by recent developments in string theory, considerable research has focused on extending the
general relativity to higher-dimensional spacetimes, which appears to be essential for a unified 
framework of fundamental interactions.
In this context, the incorporation of additional geometric structures in the gravitational action—such
as Lovelock terms \cite{Lovelock}, or brane-like components \cite{B1, B2, B3, B4} has enriched 
the landscape of theoretical model. Lovelock gravity, in particular, represents a natural generalization
of general relativity to spacetimes with dimensions $n\geq 5$. The action imposed in this theory aligns
with string theory–inspired corrections to the Einstein-Hilbert action \cite%
{string}. A distinctive and compelling feature of Lovelock gravity is that, despite involving higher-order
curvature terms,the field equations continue to remain second order, irrespective of
the fact that they are accompanied by higher-order polynomials of curvature
tensors. This traces back to the topological interpretation of each Lovelock
term as the dimensionally continued Euler densities. Therefore, there appear
no ghost degrees of freedom at the linearized level \cite{Zwi1985, Zum1986}
and the Lovelock gravity is a classically well-posed gravitational theory.
Extensive research has been conducted on exact black hole solutions in Lovelock gravity, 
particularly those featuring horizons of constant curvature and incorporating second- 
\cite{GB} and third- \cite{Lovelockex} order curvature corrections.

Lovelock gravity includes a diverse range of black hole solutions, 
notably those with horizons described by Einstein manifolds that exhibit non-constant curvature.
Specially, one can substitute the usual $(n-2)$ sphere of the horizon
geometry with an $(n-2)-$dimensional Einstein manifold. 
The first explicit example of a compact inhomogeneous Einstein metric
in four dimensions was constructed by Page \cite{Page} and generalized
to higher dimensions \cite{Hashimoto}. Bohm constructed an infinite family
of inhomogeneous metrics with positive scalar curvature on products of
spheres \cite{Bohm} and examples in higher-dimensional spacetimes has been
worked on \cite{Gibbons1, LuPa, Gauntlett, Gibbons2, Gibbons3, Gibbons4}.
But the situation is different when higher order curvature terms such as the
Lovelock terms, are introduced. In general relativity, Einstein's equations
only involve the Ricci tensor. It was shown in \cite{Dotti} 
that in the presence of the Gauss-Bonnet term, direct contribution of
the Riemann tensor and subsequently, appearance of the Weyl tensor in the
field equation, leads to new solution that changes the
properties of the spacetime. The properties of these solutions have been examined in detail in Refs. \cite{Dotti2, Maeda1}. 
In third-order Lovelock gravity, the inclusion of higher-order curvature terms imposes two distinct 
algebraic constraints on the Weyl tensor \cite{Far1}. For black holes whose base manifolds possess 
nonconstant curvature, Ref. \cite{Ray} derived general tensorial conditions imposed on the horizons 
by Lovelock field equations of arbitrary order, and further demonstrated that these conditions are 
equivalent to those expressed in terms of tensors constructed from the conformal Weyl tensor. 
The vacuum behavior and geometric properties of such black holes were subsequently analyzed in 
Ref. \cite{Ohashi}. Moreover, higher-curvature corrections beyond the Einstein–Hilbert action 
are known to introduce novel features in black-hole structure, influencing both their thermodynamic 
properties and causal characteristics \cite{Far2, Ali1, Ali2}.

In this work, we aim to construct black hole solutions with Einstein horizons
in the context of gravity coupled to electromagnetic fields. This class of models
is particularly compelling in the ongoing pursuit of regular black hole solutions. 
It is well known that the Reissner–Nordstr\"{o}m solutions of the Einstein–Maxwell
equations describe electrically or magnetically charged black holes. Similar to their
uncharged counterparts, these objects undergo Hawking radiation. However, in contrast
to neutral black holes, charged black holes are not expected to evaporate entirely; 
rather, their evolution is ultimately arrested by the presence of conserved charges,
which act to stabilize the final state. In \cite{Wein}, theories containing
electrically charged vector mesons which admit magnetically charged black
hole solutions with rather nontrivial structure are studied. A particle with
both electric and magnetic charge is called a dyon. Since magnetic
monopoles have been predicted in various extensions of the standard model of
particle physics, the interest in the possibility of dyonic black holes has
been grown. The solution for such black holes could be derived by replacing the
sum of electric and magnetic charge in the Kerr-Newman solution instead of
the electric charge \cite{Semiz}. But obtaining magnetic black hole
solutions and their properties is more complicated because the number of the
magnetic components of the Faraday tensor grows with the spacetime
dimensions in contrast with the standard electric solution. Magnetically
charged black holes have been studied in general relativity \cite{Nair, Mald}
and Gauss-Bonnet gravity \cite{Dehghani, Yazd}. Also magnetic solutions with
nonlinear electrodynamics \cite{Hendi}, and higher derivative gauge
corrections, \cite{Kats, Anninos}, are derived. We will build our results
from the magnetic black holes with non-maximal symmetry \cite{Maeda} and will
generalize the solutions to the third order Lovelock black holes.
In particular, we aim to investigate how the charge-like parameters, arising from the
non-constancy of the horizon geometry, influence the properties of these black holes
in addition to the conventional electric and magnetic charges.  We are interested in
studying the thermodynamics and stability characteristics of the resulting solutions.

Since the pioneering works of Bekenstein \cite{Bek} and Hawking \cite{Haw}, together with
the classical insights of Christodoulou \cite{Christ} and Bardeen, \cite{Bardeen}, black-hole 
thermodynamics has posed profound challenges to our understanding of the interface between quantum mechanics
and gravity. A major development in this field was the study of phase transitions in anti–de Sitter 
(AdS) spacetime, most notably the Hawking–Page transition \cite{Haw-Page}, which later proved to play 
a central role in gauge–gravity duality via the AdS/CFT correspondence. The emergence of extended phase-space 
thermodynamics stimulated intense research interest, revealing an unexpectedly rich landscape of novel phase transitions 
and intricate phase structures in AdS black-hole systems \cite{Kub1, Kastor, Kub2, Wei, Wei2}. Further investigations of 
charged and rotating AdS black holes uncovered thermodynamic behaviour closely analogous 
to that of ordinary fluids, including Van der Waals–type criticality, swallow-tail features in 
Gibbs free-energy diagrams, and well-defined equations of state \cite{Kub3, Kub4, Kub5, Cheng}. 
Analyses based on heat capacity have shown that divergences or sign changes in the heat capacity 
signal the onset of phase transitions or thermodynamic instabilities \cite{Dav, Shen, Barg}. 
Whereas Schwarzschild black holes in asymptotically flat spacetime are always thermodynamically 
unstable due to their negative heat capacity, it is shown that charged and rotating black holes 
possess stable regions in which phase transitions can occur \cite{Hut, Rup, Avr}. Several recent studies 
offer further detailed studies on this topic \cite{BHT}. Lovelock black holes exhibit rich thermodynamic structure due to the presence of higher-curvature 
corrections to Einstein gravity. In the extended phase space, where the cosmological constant is interpreted 
as a thermodynamic pressure, Lovelock black holes may undergo Van der Waals–type first-order phase transitions 
between small and large black-hole phases, often accompanied by non-monotonic behavior of the Hawking temperature 
and the appearance of critical points \cite{Wei3, Kub6, Bai, Hu}. In third-order Lovelock gravity, multiple 
physical critical points may arise for suitable choices of the second- and third-order Lovelock couplings 
depending on the spacetime dimension \cite{Giac, Dol, Deh1, Mo, Wang, Xu, Alkac, Zhang}. 
When black-hole horizons possess nonconstant curvature, the intrinsic geometry of the base manifold further modifies the thermodynamic quantities, 
producing an even richer variety of phase structures and critical phenomena. Depending on the interplay between horizon geometry, 
higher-curvature interactions, and dimensionality, such black holes can display Van der Waals–type small/large transitions, 
reentrant phase transitions, or multiple critical points. Complementary studies reinforce this conclusion. In \cite{Hull1}, 
it is shown that how Lovelock couplings and horizon topology shape the conditions and
nature of Hawking-Page, small-large, and triple point phenomena of third-order Lovelock exotic
black hole solutions. In \cite{Hull2} authors show that asymptotically AdS black holes in Gauss–Bonnet
Lovelock gravity with non-constant curvature horizons exhibit triple points and generalized mass-
less or negative mass solutions. The findings of \cite{Far3} reveal distinctive critical behaviors, 
including multiple types of phase transitions and the existence of up to three critical points for Lovelock
black holes with non-maximally symmetric horizons in vacuum.

The structure of this paper is as follows. In the next section, we begin with a 
brief review of the field equations in third-order Lovelock gravity in the presence
of an electromagnetic field. We also discuss the conditions dominating Einstein 
manifold imposing by second and third order Lovelock terms in field equations. 
In Section \ref{BHS}, we derive charged black hole solutions with nonconstant-curvature
horizons by incorporating an electromagnetic field. Section \ref{Thermo} is dedicated to
the analysis of the thermodynamic properties of these solutions, with particular attention
to their stability. Finally, we conclude with a summary of our results and some closing remarks.

\section{Field Equations In Einstein-Maxwell-Lovelock Gravity}

\label{NCL}
In Lovelock gravity, the Lagrangian is constructed from a series of curvature invariants known 
as Lovelock terms. The $m$th-order Lovelock Lagrangian is given by
\begin{equation}
\mathcal{L}_{m}=\frac{1}{2^{m}}\delta _{\rho _{1}\kappa _{1}\cdots \rho
_{m}\kappa _{m}}^{\lambda _{1}\sigma _{1}\cdots \lambda _{m}\sigma
_{m}}R_{\lambda _{1}\sigma _{1}}{}^{\rho _{1}\kappa _{1}}\cdots R_{\lambda
_{m}\sigma _{m}}{}^{\rho _{m}\kappa _{m}}\ ,  \label{Lag1}
\end{equation}%
where $R_{\lambda \sigma }{}^{\rho \kappa }$ is the Riemann tensor in $n$%
-dimensions and $\delta _{\rho _{1}\kappa _{1}\cdots \rho _{m}\kappa
_{m}}^{\lambda _{1}\sigma _{1}\cdots \lambda _{m}\sigma _{m}}$ is the
generalized totally antisymmetric Kronecker delta. The total Lovelock Lagrangian in 
$n$-dimensions spacetime is then expressed as a finite sum:

\begin{equation}
L=\sum_{m=0}^{k}\alpha _{m}\mathcal{L}_{m}\ ,  \label{Lag2}
\end{equation}

where $\alpha _{m}$  are the Lovelock coupling constants.
In the presence of an electromagnetic field, the total action reads

\begin{equation}
S=\int d^{n}x\sqrt{-g}\left[ -2\Lambda +\sum_{m=1}^{k}\left\{ \alpha _{m}%
\mathcal{L}_{m}\right\} -F_{\mu \nu }F^{\mu \nu },\right]  \label{Act}
\end{equation}%
\bigskip where the Maxwell field strength, or the Faraday tensor, is given
by $F_{\mu \nu }:=\partial _{\mu }A_{\nu }-\partial _{\nu }A_{\mu }$ with $%
A^{\mu }$ being the vector potential. Variation of the action with respect to the metric
to $g_{\mu \nu },$ yields the field equations
\begin{equation}
\mathcal{G}_{\mu }{}^{\nu }=T_{\mu }{}^{\nu },  \label{FE1}
\end{equation}%
where $\mathcal{G}_{\mu }{}^{\nu }$ is Lovelock tensor defined as
\begin{equation}
\mathcal{G}_{\mu }{}^{\nu }=\Lambda \delta _{\mu }^{\nu }-\sum_{m=1}^{k}%
\frac{1}{2^{m+1}}\frac{a_{m}}{m}\delta _{\mu \rho _{1}\kappa _{1}\cdots \rho
_{m}\kappa _{m}}^{\nu \lambda _{1}\sigma _{1}\cdots \lambda _{m}\sigma
_{m}}R_{\lambda _{1}\sigma _{1}}{}^{\rho _{1}\kappa _{1}}\cdots R_{\lambda
_{m}\sigma _{m}}{}^{\rho _{m}\kappa _{m}}  \label{LT}
\end{equation}%
and $T_{\mu \nu }$ is given by
\begin{equation}
T_{\mu \nu }=F_{\mu \rho }F_{\nu }^{~\rho }-\frac{1}{4}g_{\mu \nu }\mathcal{%
F,}  \label{EMT}
\end{equation}%
\begin{equation}
\mathcal{F}:=F_{\mu \nu }F^{\mu \nu }  \label{FF}
\end{equation}%
The Maxwell equation reads
\begin{equation}
\nabla _{\nu }F^{\mu \nu }=0.  \label{Max-Eq}
\end{equation}

\bigskip

\subsection{Metric Ans\"{a}tze}
We consider an $n$-dimensional manifold $\mathcal{M}^{n}$ to be defined as follows: 
\begin{equation}
g_{\mu \nu }dx^{\mu }dx^{\nu }=g_{ab}(y)dy^{a}dy^{b}+r^{2}(y)\gamma
_{ij}(z)dz^{i}dz^{j},  \label{Metric}
\end{equation}%
which represents a warped product of a two-dimensional Riemannian
submanifold $\mathcal{M}^{2}$ given by 
\begin{equation}
ds^{2}=-f(r)dt^{2}+g(r)dr^{2}.  \label{metric1}
\end{equation}%
and an $(n-2)$-dimensional submanifold $\mathcal{K}^{(n-2)}$ with the metric 
\begin{equation}
ds^{2}=r^{2}\gamma _{ij}(z)dz^{i}dz^{j}.  \label{metric2}
\end{equation}%
Here we assume the submanifold $\mathcal{K}^{(n-2)}$ with the unit metric $%
\gamma _{ij}$ is an Einstein manifold with nonconstant curvature and volume $%
V_{n-2}$, where $i,j=2...n-1$. The Ricci scalar, Ricci tensor, and Riemann
tensor of this submanifold can be stated as

\begin{eqnarray}
\widetilde{R} &=&\kappa (n-2)(n-3),\text{ \ \ \ }  \label{Ricci Sca} \\
\text{\ \ \ \ }\widetilde{R}_{ij} &=&\kappa (n-3)\gamma _{ij},
\label{Ricci Ten} \\
\widetilde{{R}}{_{ij}}^{kl} &=&\widetilde{{C}}{_{ij}}^{kl}+\kappa ({\delta
_{i}}^{k}{\delta _{j}}^{l}-{\delta _{i}}^{l}{\delta _{j}}^{k})\text{\ },
\label{Riemm Ten}
\end{eqnarray}%
with $\kappa $ being the sectional curvature and  $\widetilde{{C}}{_{ij}}%
^{kl}$ is the Weyl tensor of $\mathcal{K}^{(n-2)}$. Hereafter we use tilde
for the tensor components of the submanifold $\mathcal{K}^{(n-2)}.$ 

For the metric (\ref{metric1}) to be a solution of field equations in third
order Lovelock theory in vacuum, it would suffice that the Weyl tensor of
the horizon satisfies the following constraints 
\begin{equation}
\sum_{kln}\widetilde{{C}}{_{ki}}^{nl}\widetilde{{C}}{_{nl}}^{kj}=\frac{1}{n}{%
\delta _{i}}^{j}\sum_{kmpq}\widetilde{{C}}{_{km}}^{pq}\widetilde{{C}}{_{pq}}%
^{km}\equiv \eta _{2}{\delta _{i}}^{j},  \label{eta2}
\end{equation}

\begin{eqnarray}
&&\sum_{klnmp}2(4\widetilde{{C}}{^{nm}}_{pk}\widetilde{{C}}{^{kl}}_{ni}%
\widetilde{{C}}{^{pj}}_{ml}-\widetilde{{C}}{^{pm}}_{ni}\widetilde{C}^{jnkl}%
\widetilde{C}_{klpm})  \nonumber \\
&=&\frac{2}{n}{\delta _{i}}^{j}\sum_{klmpqr}\left( 4\widetilde{{C}}{^{qm}}%
_{pk}\widetilde{{C}}{^{kl}}_{qr}\widetilde{{C}}{^{pr}}_{ml}-\widetilde{{C}}{%
^{pm}}_{qr}\widetilde{C}^{rqkl}\widetilde{C}_{klpm}\right)   \nonumber \\
&\equiv &\eta _{3}{\delta _{i}}^{j}.  \label{eta3}
\end{eqnarray}%
The first constraint was originally introduced by Dotti and Gleiser in \cite%
{Dotti} and the second one which is dictated by the third order Lovelock
term, is obtained in \cite{Far1}. where $\hat{\alpha}_{p}$ are defined as $%
\hat{\alpha}_{0}\equiv -2\Lambda /(n-1)(n-2)$, $\hat{\alpha}_{2}\equiv
(n-3)(n-4)\alpha _{2}$ and $\hat{\alpha}_{3}\equiv (n-3)!\alpha _{3}/(n-7)!$
for simplicity.

Making use of the definitions above, the $tt$ and $rr$ components of field
equation (\ref{LT}) reduce to 
\begin{eqnarray}
{\mathcal{G}_{t}}^{t} &=&\frac{(n-2)}{2r^{6}g^{4}}\{[r^{4}g^{2}+3\hat{\alpha}%
_{3}\hat{\eta}_{2}g^{2}+2\hat{\alpha}_{2}r^{2}(kg-1)g+3\hat{\alpha}%
_{3}(kg-1)^{2}]rg^{^{\prime }}+(kg-1)[(n-3)r^{4}g^{2}  \nonumber \\
&&+3(n-7)\hat{\alpha}_{3}\hat{\eta}_{2}g^{2}+(n-5)\hat{\alpha}%
_{2}r^{2}(kg-1)g+(n-7)\hat{\alpha}_{3}(kg-1)^{2}]g  \nonumber \\
&&+\left( (n-1)\hat{\alpha}_{0}+\frac{(n-5)\hat{\alpha}_{2}\hat{\eta}_{2}}{%
r^{4}}+\frac{(n-7)\hat{\alpha}_{3}\hat{\eta}_{3}}{r^{6}}\right) r^{6}g^{4}\},
\label{Gtt}
\end{eqnarray}%
\begin{eqnarray}
{\mathcal{G}_{r}}^{r} &=&\frac{(n-2)}{2r^{6}fg^{3}}\{[r^{4}g^{2}+3\hat{\alpha%
}_{3}\hat{\eta}_{2}g^{2}+2\hat{\alpha}_{2}r^{2}(kg-1)g+3\hat{\alpha}%
_{3}(kg-1)^{2}]rf^{\prime }-(kg-1)[(n-3)r^{4}g^{2}  \nonumber \\
&&+3(n-7)\hat{\alpha}_{3}\hat{\eta}_{2}g^{2}+(n-5)\hat{\alpha}%
_{2}r^{2}(kg-1)g+(n-7)\hat{\alpha}_{3}(kg-1)^{2}]f  \nonumber \\
&&+\left( (n-1)\hat{\alpha}_{0}+\frac{\hat{\alpha}_{2}(n-5)\hat{\eta}_{2}}{%
r^{4}}+\frac{(n-7)\hat{\alpha}_{3}\hat{\eta}}{r^{6}}\right) r^{6}g^{4}\},
\label{Grr}
\end{eqnarray}

where we have used the definition $\hat{\eta}_{2}=(n-6)!\eta _{2}/(n-2)!$
and $\hat{\eta}_{3}=(n-8)!\eta _{3}/(n-2)!$ for simplicity. It is notable to
mention that for these kinds of Einstein metrics $\hat{\eta}_{2}$ is always
positive, but $\hat{\eta}_{3}$ can be positive or negative relating to the
metric of the spacetime.

\section{Black Hole Solutions in the Presence of Electromagnetic Field In Even
Dimensions}

\label{BHS}
We assume the energy-momentum tensor to have the following form
\begin{equation}
T_{\mu \nu }dx^{\mu }dx^{\nu }=T_{ab}(y)dy^{a}dy^{b}+p(y)r^{2}(y)\gamma
_{ij}dz^{i}dz^{j},
\end{equation}%
where $p(y)$ is a scalar function. We consider the vector potential to be as
following \
\begin{equation}
A_{\mu }dx^{\mu }=A_{a}(y)dy^{a}+A_{i}(z)dz^{i},
\end{equation}%
in analogy with the spacetime ansatz (\ref{Metric}), by which we can write
Faraday tensor as
\begin{equation}
F_{\mu \nu }dx^{\mu }\wedge dx^{\nu }=F_{ab}(y)dy^{a}\wedge
dy^{b}+F_{ij}(z)dz^{i}\wedge dz^{j}.
\end{equation}

We assume $F_{ab}(y)$ and $F_{ij}(z)$ to be the corresponding electric and
magnetic components, respectively. For the magnetic component, we add the
following assumption
\begin{equation}
\gamma ^{kl}F_{ik}F_{jl}=q_{m}^{2}\gamma _{ij},
\end{equation}%
where $q_{m}$ is a constant. This condition is the consequence of the field
equations \cite{Ortaggio}. Subsequently the Maxwell invariant scalar reads
\begin{equation}
\mathcal{F}=2F_{tr}F^{tr}+\frac{(n-2)C^{2}}{r^{4}},
\end{equation}

\bigskip We consider $r^{2}\gamma _{ij}(z)dz^{i}dz^{j}$ as a $(n-2)$%
-dimensional maximally symmetric submanifold $\mathcal{K}^{(n-2)}$ with
positive curvature containing $2$-dimensional hypersurfase as $d\theta
^{2}+\sin ^{2}(\theta )d\phi ^{2}$. Knowing that the electric field is
associated with the time component, $A_{t}$ of the vector potential and the
magnetic field is associated with the angular component $A_{\varphi},$ we
can write the relations for electric and magnetic gauge field $A_{\mu }$ as
\begin{eqnarray}
A_{a} &=&\frac{Q_{e}}{(n-3)r^{(n-3)}}\delta _{a}^{t}  \notag \\
A_{i} &=&Q_{m}\cos \theta \delta _{i}^{\phi }
\end{eqnarray}%
where $Q_{e}$ is the electric and $Q_{m}$ is the magnetic charge. This
implies that the only non-vanishing components of the symmetric Maxwell tensor
are $F_{tr}$ and $F_{r\varphi}$. Accordingly, we construct black hole solutions
carrying both magnetic and electric charges by extending the standard
four-dimensional result in higher dimensions. A notable example of such a 
solution is an Einstein space formed as the product of two maximally symmetric spaces. 
So that the energy-momentum tensor (\ref{EMT}) is calculated
to be%
\begin{eqnarray}
T_{~~b}^{a} &=&[\frac{1}{2}\biggl(F_{tr}F^{tr}-\frac{(n-2)C^{2}}{2r^{4}}%
\biggl)]\delta _{~~b}^{a}  \label{Tab} \\
T_{~~j}^{i} &=&[-\frac{1}{2}\biggl(F_{tr}F^{tr}+\frac{(n-6)C^{2}}{2r^{4}}%
\biggl)]\delta _{~~j}^{i}.  \label{Tij}
\end{eqnarray}

With $C=(n-3)Q_{m\text{ ,}}$Eqs. (\ref{Tab}) and (\ref{Tij}) are calculated
for the metric (\ref{Metric})\ to be%
\begin{eqnarray}
T_{ab} &=&-(\frac{Q_{e}^{2}}{r^{2(n-2)}}+\frac{(n-2)Q_{m}^{2}}{2r^{4}})g_{ab}
\\
\widetilde{T}_{ij} &=&(\frac{Q_{e}^{2}}{r^{2n-6}}-\frac{(n-6)Q_{m}^{2}}{%
2r^{2}})\gamma _{ij},
\end{eqnarray}

The vacuum equation $\mathcal{G}_{t}^{t}-\mathcal{G}%
_{r}^{r}=0$ implies that $d(fg)/dr=0,$ and therefore one can take $%
g(r)=1/f(r)$ by rescaling the time coordinate $t$. The $tt$ component of field equation (\ref{FE1}) is then simplified to be

\begin{eqnarray}
&&\frac{(n-2)}{2r^{6}}\{[r^{5}+2\widehat{\alpha }_{2}r^{3}(\kappa -f)+3%
\widehat{\alpha }_{3}r(\widehat{\eta }_{2}+(\kappa -f)^{2})]f^{^{\prime
}}-(\kappa -f)[(n-3)r^{4}  \notag \\
&&+(n-5)\widehat{\alpha }_{2}r^{2}(\kappa -f)+(n-7)\widehat{\alpha }_{3}(3%
\widehat{\eta }_{2}+(\kappa -f)^{2})]  \notag \\
&&-\left( (n-1)\widehat{\alpha }_{0}+\frac{(n-5)\widehat{\alpha }_{2}%
\widehat{\eta }_{2}}{r^{4}}+\frac{(n-7)\widehat{\alpha }_{3}\widehat{\eta }%
_{3}}{r^{6}}\right) r^{6}\}={\mathcal{G}_{t}}^{t}=\frac{Q_{e}^{2}}{r^{2n-10}}%
+\frac{Q_{m}^{2}}{2}(n-2)r^{2}
\end{eqnarray}%
It is important to note that the class of solutions under consideration exists
only in even-dimensional spacetimes, as dictated by the structure of the field equations, from which we obtain:
\begin{equation}
\gamma _{ij}F^{2}=(n-2)F_{ik}F_{jl}\gamma ^{kl}, \label{EqF}
\end{equation} 

Taking the determinant of (\ref{EqF}), we obtain $(F^{2})^{n-2}\gamma
^{2}=(2-n)^{n-2}(\det F_{ij})^{2},$ knowing that $F_{ij}$ is an
antisymmetric matrix and $\det F_{ij}=\det (-F_{ij})=(-1)^{n-2}\det F_{ij},$
we see that $F^{2}$\ is zero in any odd dimensins. Introducing

\begin{equation}
\psi (r)=\frac{\kappa -f(r)}{r^{2}},  \label{Psi}
\end{equation}
and integrating $\int r^{n-2}\mathcal{G}_{t}^{t}dr$, one obtains

\begin{equation}
\left( 1+\frac{3\widehat{\alpha }_{3}\widehat{\eta }_{2}}{r^{4}}\right) \psi
+\widehat{\alpha }_{2}\psi ^{2}+\widehat{\alpha }_{3}\psi ^{3}+\widehat{%
\alpha }_{0}+\frac{\widehat{\alpha }_{2}\widehat{\eta }_{2}}{r^{4}}+\frac{%
\widehat{\alpha }_{3}\widehat{\eta }_{3}}{r^{6}}-\frac{m}{r^{n-1}}+\frac{%
Q_{e}^{2}}{(n-3)r^{2(n-2)}}+\frac{Q_{m}^{2}}{(n-5)r^{4}}=0, \label{MEE}
\end{equation} 
One of the real solutions to this equation may be written as:
\begin{eqnarray}
\psi (r) &=&-\frac{\widehat{\alpha } _{2}r^{2}}{3\widehat{\alpha }_{3}}\left\{ 1-\left(
j(r)\pm \sqrt{\gamma +j^{2}(r)}\right) ^{1/3}+\gamma ^{1/3}\left( j(r)\pm
\sqrt{\gamma +j^{2}(r)}\right) ^{-1/3}\right\} ,  \notag \\
j(r) &=&-1+\frac{9\widehat{\alpha }_{3}}{2\widehat{\alpha }_{2}^{2}}-\frac{27%
\widehat{\alpha }_{3}^{2}}{2\widehat{\alpha }_{2}^{3}}\left( \widehat{\alpha }_{0}-%
\frac{m}{r^{n-1}}+\frac{\widehat{\alpha }_{3}\widehat{\eta }_{3}}{r^{6}}+%
\frac{Q_{e}^{2}}{(n-3)r^{2(n-2)}}+\frac{Q_{m}^{2}}{(n-5)r^{4}}\right) ,\text{
\ }  \notag \\
\text{\ \ \ \ }\gamma &=&\left( -1+\frac{3\widehat{\alpha }_{3}}{\widehat{%
\alpha }_{2}^{2}}+\frac{9\widehat{\alpha }_{3}^{2}\widehat{\eta }_{2}}{%
\widehat{\alpha }_{2}^{2}r^{4}}\right) ^{3},  \label{fstat}
\end{eqnarray}%
The derived solution represents the most general form of a charged black hole
in third-order Lovelock gravity in even-dimensional spacetimes, under the influence
of an electromagnetic field.Since the constant $\widehat{\eta }_{2}$ and $\widehat{\eta }%
_{3} $ are defined on the $(n-2)$-dimensional boundary, $n$ must be at least eight for
the non-constancy of the horizon curvature to contribute at third order in Lovelock gravity.
Thus One may note that solution (\ref{fstat}) reduces to the algebraic equation of
Lovelock gravity for charged solution with constant curvature horizon when $%
\widehat{\eta }_{2}=\widehat{\eta }_{3}=0$.
An interesting feature of Eq. (\ref{MEE}) is that the term involving the magnetic charge appears 
in the same form as the term containing the charge-like parameter $\widehat{\eta }_{2}$ which originates
from the second-order Lovelock contributions in the case of a metric with a nonconstant curvature horizon.
This resemblance allows for an interpretation of $\widehat{\eta }_{2}$ as an effective magnetic charge parameter. 
Furthermore, we observe that the dominant behavior of the metric function near $r=0$
as inferred from Eq. (\ref{MEE}) is governed by
\begin{equation}
f(r)\simeq (\frac{Q_{e}^{2}}{(n-3)\widehat{\alpha }_{3}r^{2n-10}})^{1/3},
\end{equation}

It is seen that central singularity for this solution is
timelike in contrast with the uncharged solution which possesses a
spacelike singularity

\section{Thermal stability}
\label{Thermo}
The surface gravity on the Killing horizon is $(1/2)(df/dr)|_{r=r_{h}}$,
from which the temperature of the horizon $T$ could be written as

\begin{equation}
T=\frac{(n-1)r_{h}^{6}\widehat{\alpha }_{0}+(n-3)\kappa r_{h}^{4}+(n-5)%
\widehat{\alpha }_{2}(\widehat{\eta }_{2}+\kappa ^{2})r_{h}^{2}+(n-7)%
\widehat{\alpha }_{3}(\widehat{\eta }_{3}+3\kappa \widehat{\eta }_{2}+\kappa
^{3})-Q_{e}^{2}r_{h}^{10-2n}+Q_{M}^{2}r_{h}^{2}}{4\pi r_{h}[r_{h}^{4}+2\kappa
\widehat{\alpha }_{2}r_{h}^{2}+3\widehat{\alpha }_{3}(\widehat{\eta }%
_{2}+\kappa ^{2})]},  \label{Temp}
\end{equation}%
where $r_{h}$ is the radius of the outer horizon. On the other hand, the
entropy on the Killing horizon is calculated using the Wald prescription
which is applicable for any black hole solution of which the event horizon
is a killing horizon \cite{Wald}. The Wald entropy is defined by the
following integral performed on $(n-2)$-dimensional spacelike bifurcation
surface

\begin{equation}
S=-2\pi \oint d^{n-2}x\sqrt{h}Y,\text{ \ \ \ \ \ }Y=Y^{abcd}\widehat{%
\varepsilon }_{ab}\widehat{\varepsilon }_{cd},\text{\ \ \ \ \ \ }Y^{abcd}=%
\frac{\partial \mathcal{L}}{\partial R_{abcd}}  \label{entropy}
\end{equation}%
in which $\mathcal{L}$ is the Lagrangian and $\widehat{\varepsilon }_{ab}$
is the binormal to the horizon. As we mentioned before, $\mathcal{L}_{1},$ $%
\mathcal{L}_{2}\mathcal{\ }$and $\mathcal{L}_{3},$ are Einstein,
Gauss-Bonnet and third order Lovelock Lagrangians respectively, from which
we obtain $Y_{1},$ $Y_{2}$ and $Y_{3.}$ Following the given description, $%
Y_{1}$ and $Y_{2}$ \ and $Y_{3}$ are calculated to be

\begin{equation}
Y_{1}=-\frac{1}{8\pi }  \label{Ein-entropy}
\end{equation}

\begin{equation}
Y_{2}=-\frac{\widehat{\alpha }_{2}}{4\pi }%
[R-2(R_{t}^{t}+R_{r}^{r})+2R_{tr}^{tr}]  \label{Gauss-entropy}
\end{equation}

\begin{eqnarray}
Y_{3} &=&-\frac{3\widehat{\alpha }_{3}}{4\pi }\{-12({R^{tm}}_{tn}{R^{rn}}%
_{rm}-{R^{tm}}_{rn}{{R^{r}}_{mt}}^{n})+12R^{trmn}R_{trmn}-24[{R^{tr}}_{tm}{%
R_{r}}^{m}-{R^{tr}}_{rm}{R_{t}}^{m}  \notag \\
&&+\frac{1}{4}(R_{mnpr}R^{mnpr}+R_{mnpt}R^{mnpt})]+3(2R{R^{tr}}_{tr}+\frac{1%
}{2}R_{mnpq}R^{mnpq})  \notag \\
&&+12({R^{t}}_{t}{R^{r}}_{r}-{R^{t}}_{r}{R^{r}}_{t}+{R^{r}}_{mrn}R^{mn}+{%
R^{t}}_{mtn}R^{mn})+12(R^{rm}R_{rm}+R^{tm}R_{tm})  \notag \\
&&-6[R_{mn}R^{mn}+R({R^{r}}_{r}+{R^{t}}_{t})]+\frac{3}{2}R^{2}\}.
\label{Lov-entropy}
\end{eqnarray}%
Substituting in Eq. (\ref{entropy}) one calculates the entropy to be

\begin{equation}
S=-2\pi \{Y_{1}+Y_{2}+Y_{3}\}=\frac{r_{h}^{n-2}}{4}\left\{ 1+\frac{2\kappa
\widehat{\alpha }_{2}(n-2)}{r_{h}^{2}(n-4)}+\frac{3\widehat{\alpha }%
_{3}(n-2)(\widehat{\eta }_{2}+\kappa ^{2})}{r_{h}^{4}(n-6)}\right\} .
\label{Entro}
\end{equation}

Also we obtain the relation for the mass density, from Eq. (\ref{MEE}),
which admits the relation below

\begin{eqnarray}
M &=&\frac{(n-2)m}{16\pi }=\frac{(n-2)}{16\pi }\{\widehat{\alpha }%
_{0}r_{h}^{n-1}+\kappa r_{h}^{n-3}+\widehat{\alpha }_{2}[\kappa ^{2}+\widehat{\eta }%
_{2}]r_{h}^{n-5}+\widehat{\alpha }_{3}[\kappa ^{3}+3\widehat{\eta }_{2}\kappa +%
\widehat{\eta }_{3}]r_{h}^{n-7}  \notag \\
&&+\frac{Q_{E}^{2}}{(n-3)r_{h}^{n-3}}+\frac{Q_{M}^{2}r_{h}^{n-5}}{(n-5)}\}.
\label{ADM}
\end{eqnarray}

Analyzing the behavior of the entropy $S(M, q_{_E}, q_{_M},..)$ under small variations of the
thermodynamic coordinates around equilibrium provides valuable insight into the thermal stability
of a system.  Irrespective of the chosen ensemble, if $S(M, q_{_E}, q_{_M},..)$ is a convex function
of the extensive variables—or equivalently, if its Legendre transform is a concave function of the 
corresponding intensive variables—the system is thermodynamically stable. Similarly, when the energy
$M(S, q_{_E}, q_{_M},...)$ is a convex function of its arguments, the system exhibits thermal stability.
Consequently, the local stability analysis can, in principle, be performed by evaluating the determinant
of the Hessian matrix $H$ constructed with respect to the extensive variables. It is important to note, 
however, that the number of relevant thermodynamic variables depends on the ensemble considered.
\subsection{Stability in the grand canonical ensemble}
\begin{figure}[!htb]
	\centering \subfigure[{\, $q_{_E}=2$}] {%
		\includegraphics[scale=0.35
		]{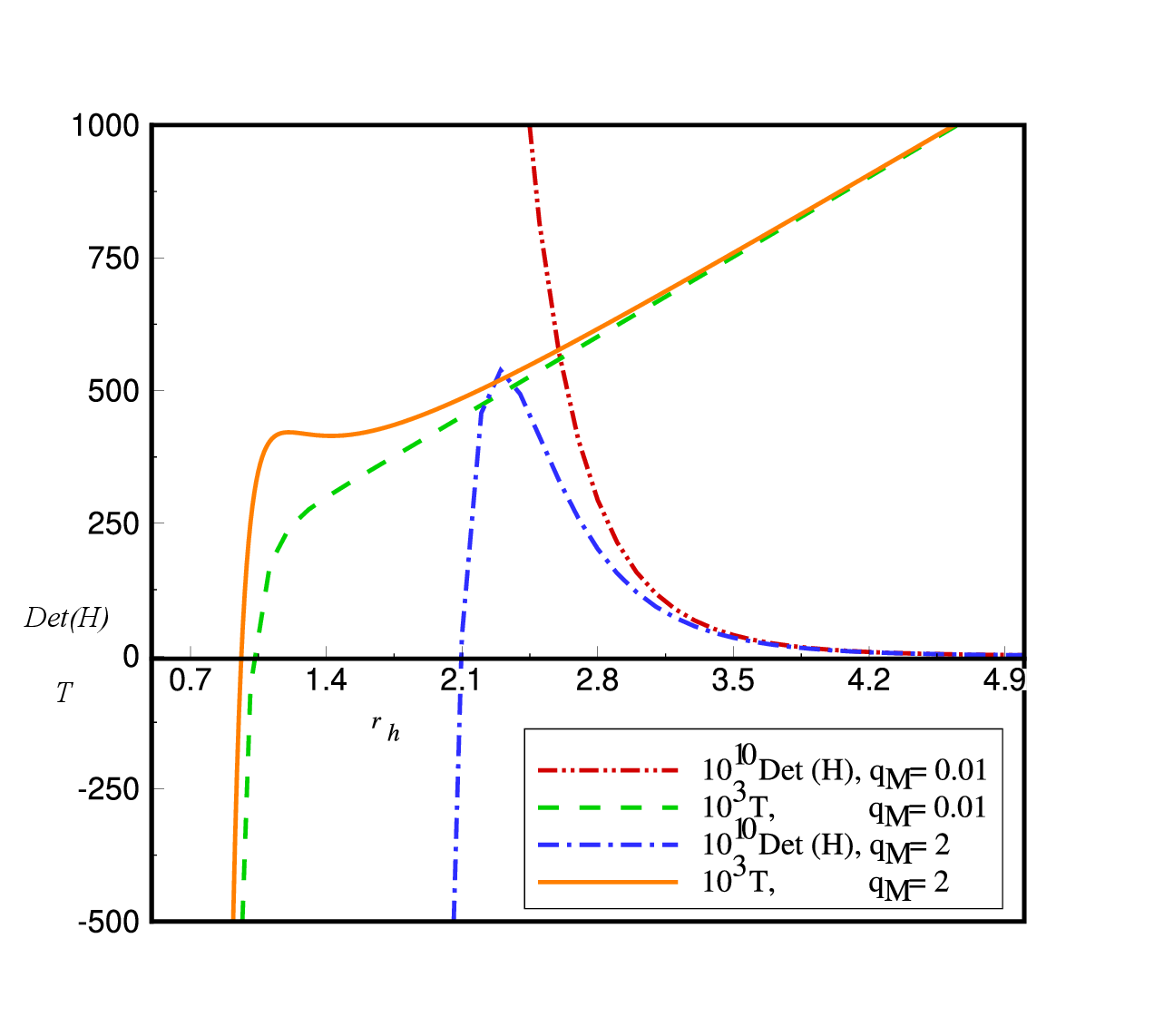}\label{fig1}} \hspace*{.07cm}
	\subfigure[{ \,$q_{_M}=1$}]{\includegraphics[scale=0.35]{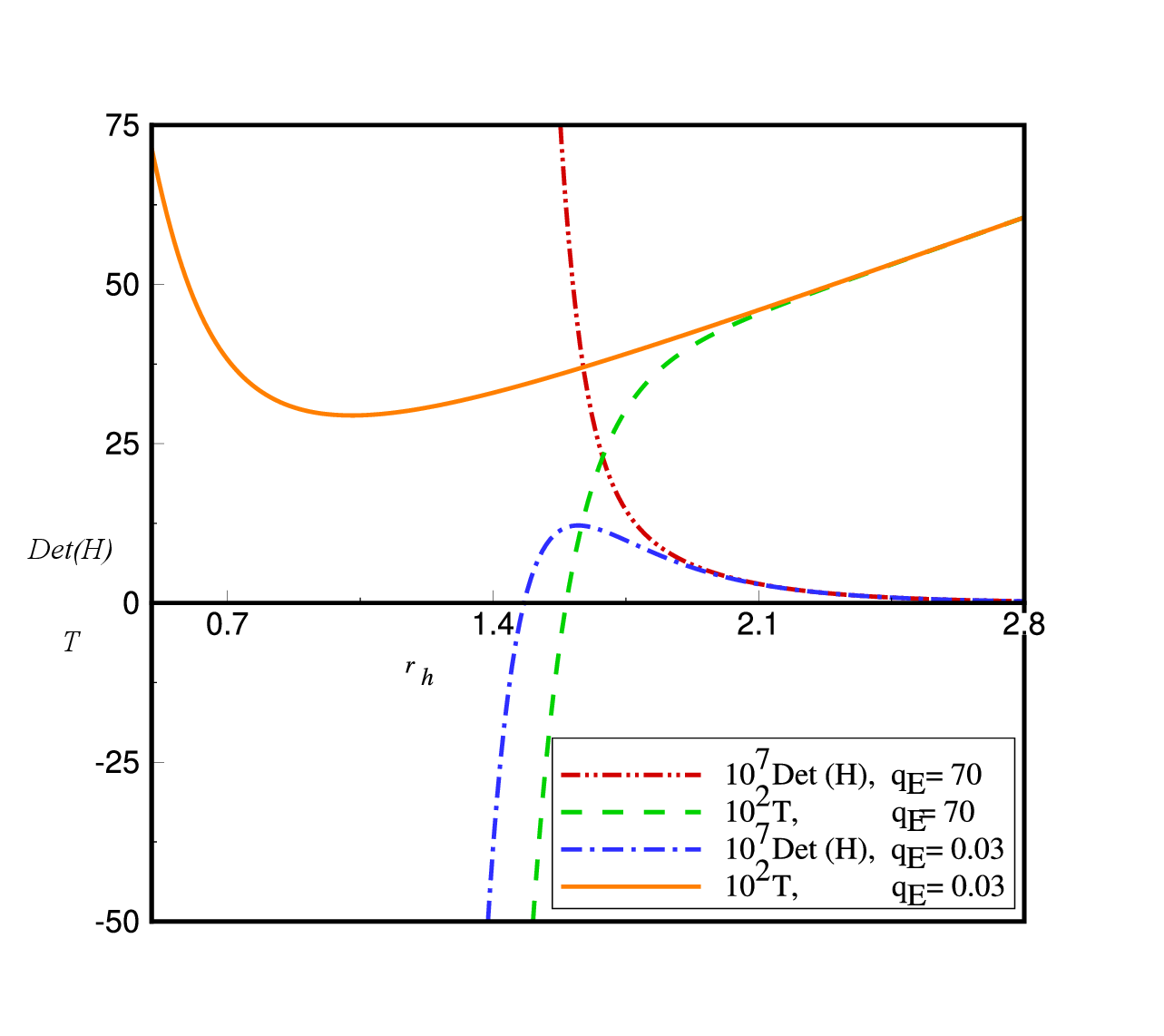}\label{fig2}}
	\caption{
		Behavior of $Det(H)$ with respect to $r_h$ for $k=0$, $n=10$, $\hat{\protect{\alpha_2}}=1$, $\hat{\protect{\alpha_{0}}}=0.3$, $\hat{\protect{\alpha_{3}}}=0.8$, $\hat{\protect\eta}_2=2$ and $\hat{\protect\eta}_3=-0.03$}
	\label{hes}
\end{figure}

This subsection is dedicated to thermal stability in the grand canonical ensemble. Our
theory contains three extensive variables $X_i$; The best way to check the
stable range of our solutions is to work in the grand canonical ensemble
\cite{Hes}. The local stability can be carried out by finding the
determinant of the Hessian matrix $(H)$. One can define $H$ as
\begin{equation}
H^{^M}_{_{X_i X_j}}=\frac{\partial^2M}{\partial X_i\partial X_j}
\end{equation}
We may select $X_i=(S,q_{_M},q_{_E})$, so the matrix components are
\begin{eqnarray}
&&H_{11}=\left(\frac{d^2M}{d^2S}\right)_{q_{_M},q_{_E}}=\left(\frac{dT}{dS}%
\right)_{q_{_M},q_{_E}}=\left(\frac{dT/dr_h}{dS/dr_h}\right)_{q_{_M},q_{_E}}
\notag \\[10pt]
&&H_{12}=H_{21}=\left(\frac{d^2M}{dSdq_{_E}}\right)_{q_{_E}}=\left(\frac{dT}{%
dq_{_E}}\right)_{S,q_{_M}}  \notag \\[10pt]
&&H_{13}=H_{31}=\left(\frac{d^2M}{dSdq_{_M}}\right)_{q_{_M}}=\left(\frac{dT}{%
dq_{_M}}\right)_{S,q_{_E}}  \notag \\[10pt]
&&H_{23}=H_{32}=\left(\frac{d^2M}{dq_{_E}dq_{_M}}\right)_{S}=0  \notag \\%
[10pt]
&& H_{22}=\left(\frac{d^2M}{d^2q_{_M}}\right)_{S,q_{_E}}  \notag \\[10pt]
&& H_{33}=\left(\frac{d^2M}{d^2q_{_E}}\right)_{S,q_{_M}}
\end{eqnarray}
It is analytically challenging to compute the determinant of the Hessian matrix, [$Det(H)$] 
so we employ graphical methods to analyze stability. For the sake of brevity, the explicit 
form of $Det(H)$ is not presented here. A positive determinant is sufficient to guarantee 
thermal stability in the grand canonical ensemble. Additionally, the system must have a 
non-negative temperature to remain stable. Figure \ref{hes} simultaneously examines these 
two conditions. For specific values of the metric parameters, instability arises in the 
case of small black holes. The plots indicate the existence of a lower bound for the 
event horizon radius  ($r_h^{min}$), based on the $Det(H)$ curves.
The solution is stable when $r_h>r_h^{m.in}$. This stability threshold may be altered by 
variations in the electric or magnetic charge parameters—specifically, increasing the 
electric charge or decreasing the magnetic charge may eliminate this lower bound.
It is notable that the temperature curves have a lower limit too, but as we mentioned before,
negative temperature values are not physically acceptable; therefore, the horizon radius must
be greater than the point at which the temperature becomes zero. Notably, the location of this
point is influenced by the system’s parameters. As shown in Figure (\ref{hes}), variations in the metric
parameters can lead to a stable black hole configuration without imposing any lower bound on the horizon radius.
\begin{figure}
	\includegraphics[scale=0.5]{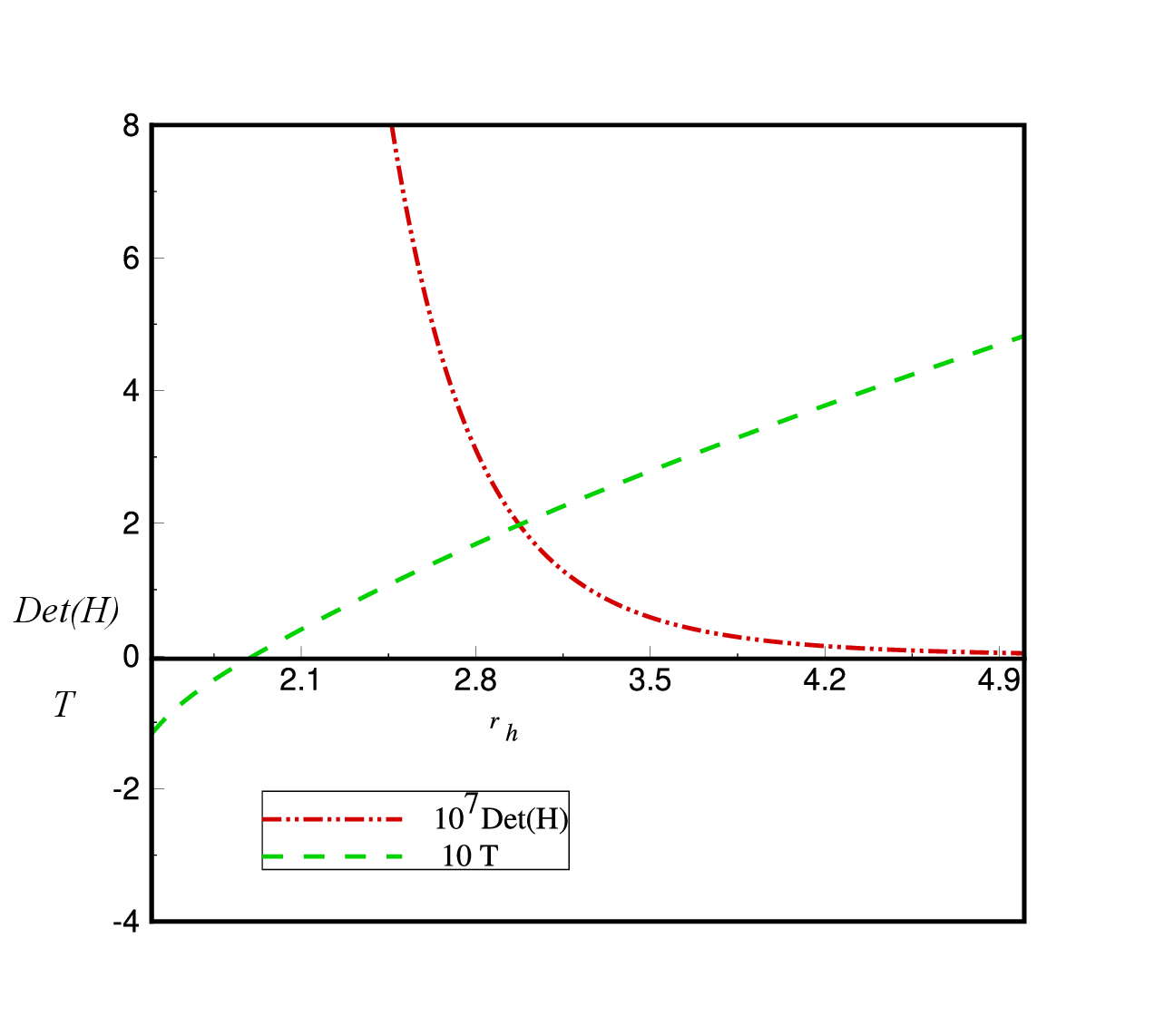}\caption{Behavior of $Det(H)$ and $T$ with respect to $r_h$ for $n=8$, $k=-1$, $\hat{\protect{\alpha_2}}=0.2$, $\hat{\protect{\alpha_{0}}}=0.2$, $\hat{\protect{\alpha_{3}}}=4$, $\hat{\protect\eta}_2=4$, $\hat{\protect\eta}_3=-0.05$, $q_{_E}=10$ and $q_{_M}=0.1$ }\label{det}
\end{figure}

Using Fig. \ref{det} one can find out that changing in metric parameters may cause a stable black hole without any restriction in horizon radius.
\subsection{Stability in the canonical ensemble}
\begin{figure}
	\includegraphics[scale=0.5]{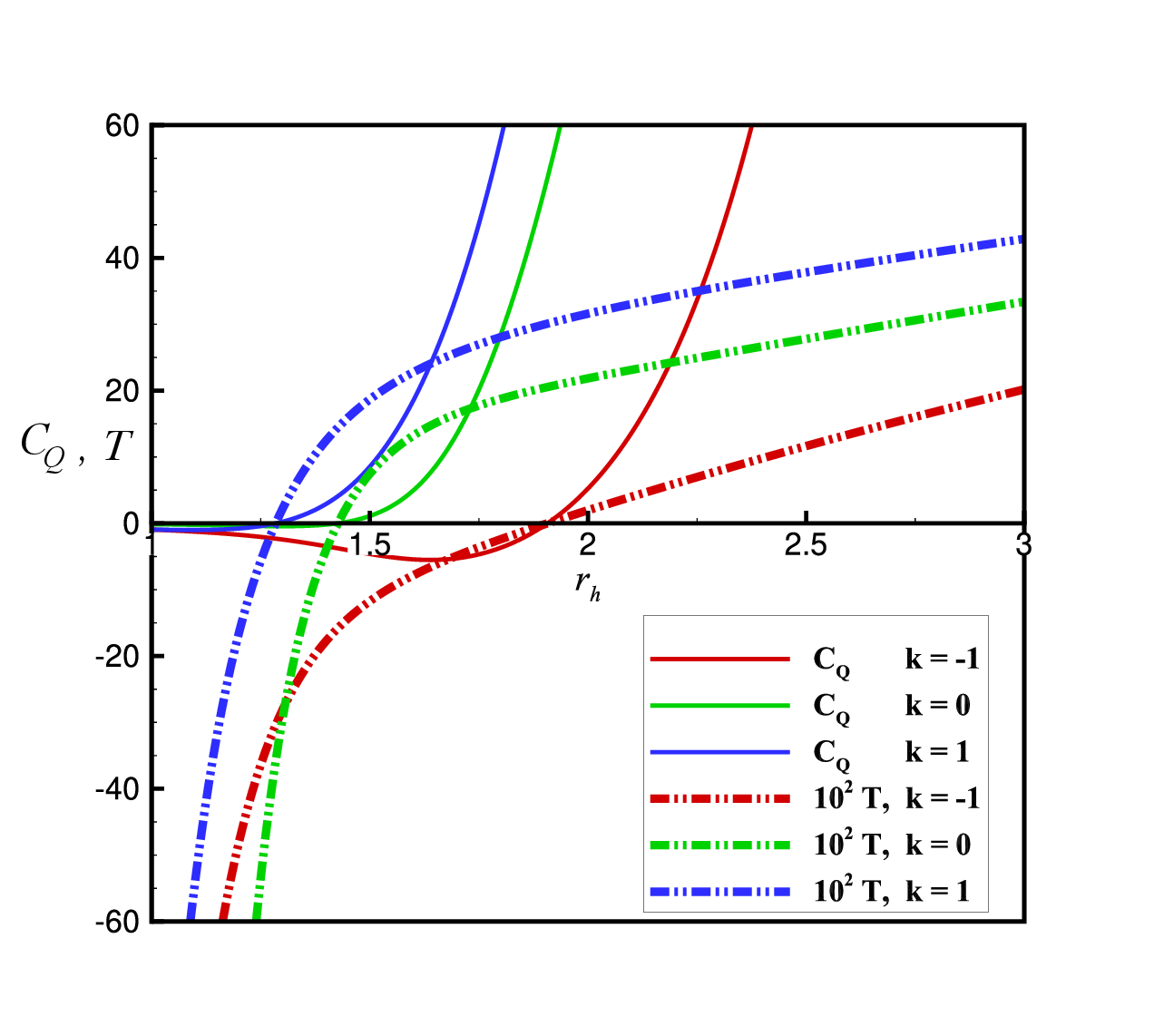}\caption{Behavior of $C_Q$ and $T$ with respect to $r_h$ for $n=8$, $\hat{\protect{\alpha_2}}=0.2$, $\hat{\protect{\alpha_{0}}}=0.2$, $\hat{\protect{\alpha_{3}}}=4$, $\hat{\protect\eta}_2=4$, $\hat{\protect\eta}_3=-0.05$, $q_{_E}=10$ and $q_{_M}=0.1$ }\label{c3}
\end{figure}

\begin{figure}
	\includegraphics[scale=0.5]{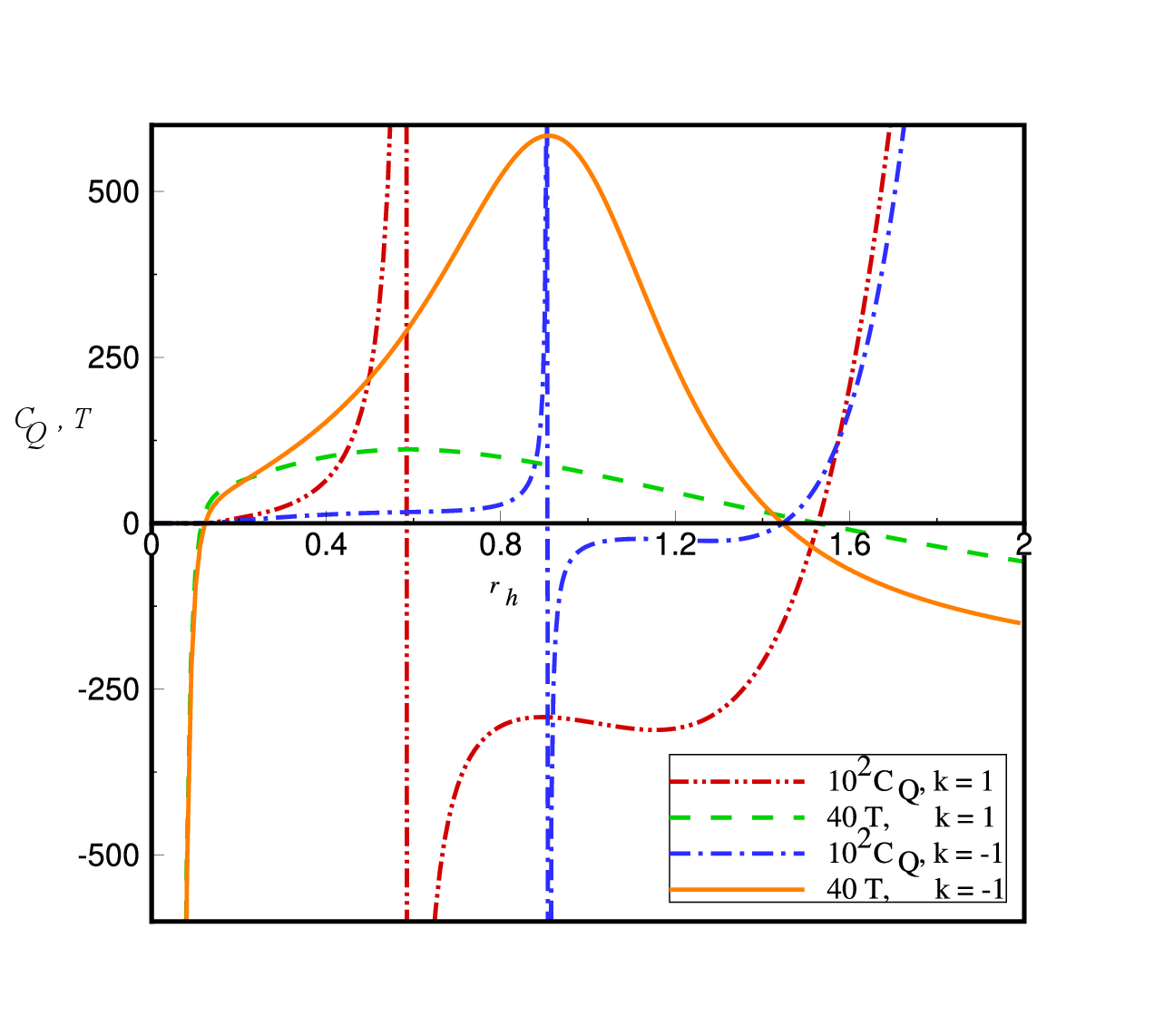}\caption{Behavior of $C_Q$ and $T$ with respect to $r_h$ for $n=8$, $\hat{\protect{\alpha_2}}=0.8$, $\hat{\protect{\alpha_{0}}}=-3$, $\hat{\protect{\alpha_{3}}}=1.05$, $\hat{\protect\eta}_2=1$, $\hat{\protect\eta}_3=-0.005$, $q_{_E}=0.002$ and $q_{_M}=10$ }\label{c1}
\end{figure}

\begin{figure}
	\includegraphics[scale=0.5]{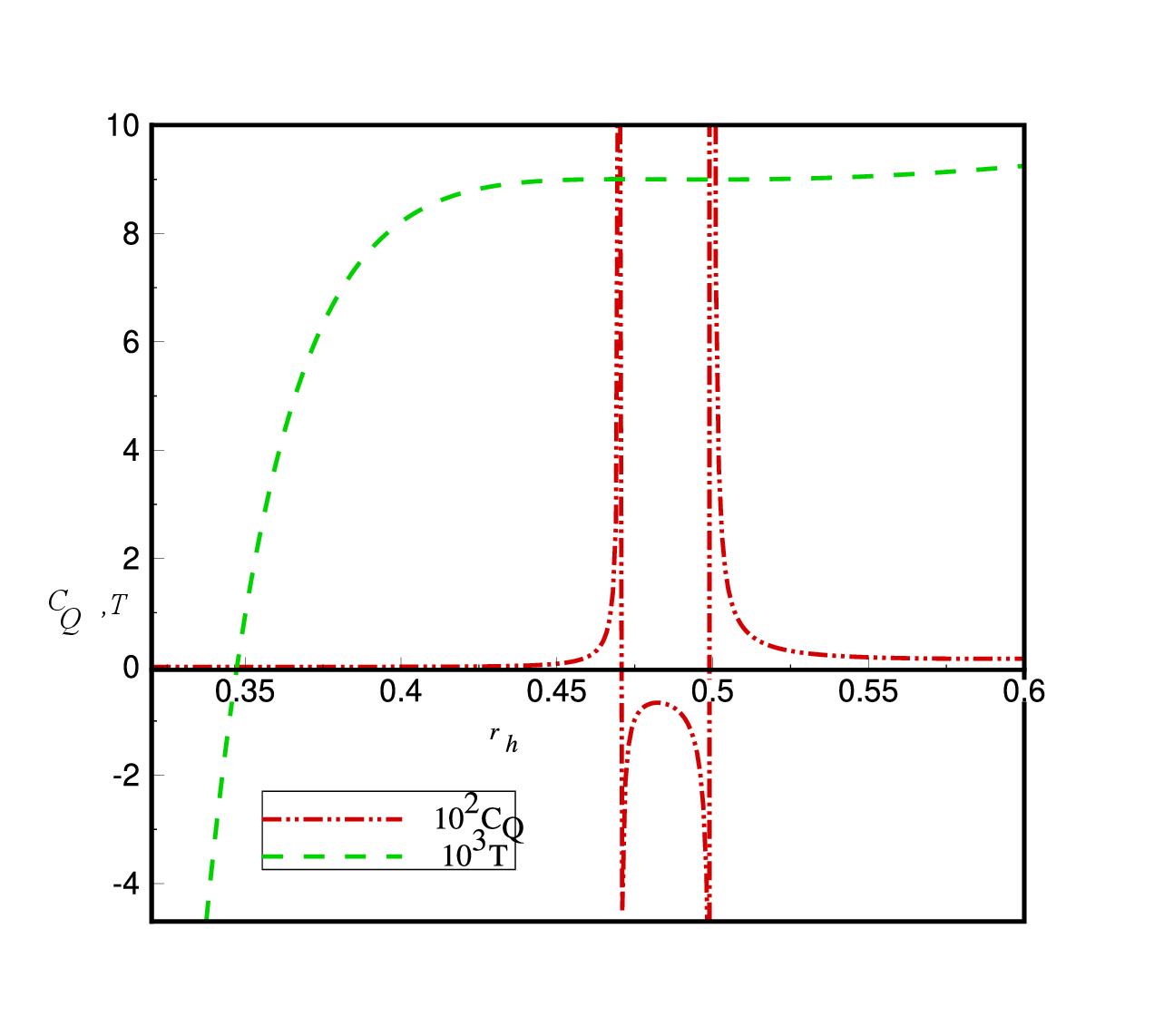}\caption{Behavior of $C_Q$ and $T$ with respect to $r_h$ for $k=0$, $n=8$, $\hat{\protect{\alpha_2}}=2$, $\hat{\protect{\alpha_{0}}}=0.4$, $\hat{\protect{\alpha_{3}}}=0.3$, $\hat{\protect\eta}_2=4$, $\hat{\protect\eta}_3=-3$, $q_{_E}=0.0013$ and  $q_{_M}=0.01$ }\label{c2}
\end{figure}

In the canonical ensemble both electric and magnetic charges are considered as fixed parameters, so the positivity of the
heat capacity $C_{q_{_M},q_{_E}}$ guarantees local stability. The mentioned function is defined as
\begin{equation}\label{c}
C_{Q}=T\left(\frac{\partial S}{\partial T}\right)=T\left(\frac{dS/dr_h}{dT/dr_h}\right)_{q_{_M},q_{_E}}
\end{equation}
here $Q=\{q_{_M},q_{_E}\}$. The temperature is not an explicit function
of entropy, so we use chain derivative in the last part of Eq. (\ref{c}). Because of complexity we use figures [Fig. \ref{c3} - Fig. \ref{c2}] to study the heat capacity function.

In order to have stable thermodynamic system in canonical ensemble, heat capacity should be positive. Unstable system experiences phase transition to change its condition. Regarding the behavior of $C_Q$, one may class phase transition to two types, type one is related to the zero value of it and type two comes from its divergence. We must be careful about zero values of $T$.
In this situation heat capacity is zero too and negative temperature is not physical, thus phase transition occurs between physical and unphysical phases to create a physical system \cite{unphy}.  Fig. \ref{c3} shows this case. Depending on the metric parameters and regardless of type of topology, $C_Q$ is positive when $T>0$, so for this class of parameters, the solution is stable for all horizon radii.

Fig. \ref{c1} displays a divergence in the $C_Q$ curves while $T>0$ so it sounds a phase transition.  Important note that it happens between physical and unphysical phase because of the negative value of heat capacity for large horizon radii. We can result that for special values of parameters there is an upper limit for $r_h^{max}$ and the black hole with $r_h>r_h^{max}$ does not exists. Note that another restriction comes from the temperature. A lower limit ($r_{h_T}^{min}$) for horizon should be considered that shows positive values of $T$. As a result, for this class of parameters, the allowed horizon radii should be selected in the $r_{h_T}^{min}<r_h<r_h^{max}$ interval. This interval expanded when $k$ decreases.

Another type of divergency is observed in Fig. \ref{c2} in the presence of positive temperature. This transition is between two physical phases which shows not allowed range of horizon because of negative $C_Q$. This type of phase transition is called second order phase transition (two divergence points). The behavior of this second order phase transition is roughly the same as that of van der Waals fluid. It divides the horizon radii region into three parts. Both the large radius region and the small radius region are thermodynamically stable with positive specific heat, while the medium radius region is unstable with negative specific heat. So there is a phase transition which takes place
between small black hole and large black hole. 
 \color{black}

\section{Concluding Remarks}
In this study, we have constructed the most general class of charged black hole solutions in third-order Lovelock gravity within even-dimensional spacetimes, under the influence of an electromagnetic field. We considered the spacetimes as the cross product of a Lorentzian manifold and a space with a nonconstant-curvature horizon. The inclusion of higher-order Lovelock terms introduces two chargelike parameters, which become dynamically relevant for spacetime dimensions is eight or greater. The solution smoothly reduces to the well-known charged Lovelock black hole with constant-curvature horizons when these parameters vanish. A notable feature of the field equations is that the term associated with the magnetic charge appears in the same form as the second-order Lovelock contribution $\widehat{\eta}_{2}$, which arises due to the nonconstant curvature of the horizon. Near the origin, the metric reveals a timelike singularity for electrically charged black holes, in contrast to the spacelike singularity in the uncharged case.

We derived the relevant thermodynamic quantities—including the Hawking temperature, Wald entropy, and mass density—and conducted a detailed thermodynamic analysis to examine the stability of these black hole solutions in both the grand canonical and canonical ensembles. In the grand canonical ensemble, where both electric and magnetic charges vary, local thermal stability is determined by the positivity of the Hessian determinant of the entropy and the temperature. Our graphical analysis reveals a lower bound on the event horizon radius for stability, which may be lifted by increasing the electric charge or decreasing the magnetic charge.
In the canonical ensemble, where the charges are held fixed, stability is governed by the sign of the specific heat. We identified two types of phase transitions: first-order transitions associated with the vanishing of the heat capacity, and second-order transitions marked by divergences. In particular, the second-order phase transition mimics the behavior of a van der Waals fluid, dividing the range of horizon radii into three regions: small and large black holes are thermodynamically stable, while intermediate-size black holes are unstable due to negative heat capacity. Depending on the metric parameters, physical consistency requires the horizon radius to lie within a finite interva $r_{h_T}^{min}<r_h<r_h^{max}$, where both temperature and heat capacity remain positive.

These findings highlight a rich and intricate phase structure shaped by the interplay between higher-curvature corrections and electromagnetic fields. The existence and thermal stability of black holes in third-order Lovelock gravity are highly sensitive to the geometric and thermodynamic parameters, offering deeper insight into gravitational dynamics and phase behavior in higher-dimensional theories.

%%%%%%%%%%%%%%%%%%%%%%%%%%%%%%%%%%%%%%%%%%%%%%%%%%%%%%%%%%%%%%%%

\end{document}